# Review on State-of-the-art Energy Systems in the Arctic

# (Pre-print)

Bilal Babar*, Sabrina Sartori*

*Department of Technology Systems, University of Oslo, Gunnar Randers vei 19, Kjeller, 2027, Norway

bilal.babar@its.uio.no

## Abstract

The Arctic regions remain heavily dependent on fossil fuels for energy generation. At the same time, the Arctic is warming at a rate considerably faster than the global average, increasing the need for low-carbon and climate-resilient energy systems. This review documents the current energy systems and assesses state-of-the-art energy solutions applicable to Arctic and cold-climate regions, as well as the future direction of these emerging energy systems, and analyses the role of energy storage, heating requirements, and advanced energy solutions. A total of 88 research articles were systematically reviewed. The reviewed studies indicate the potential for wind and solar to reduce dependence on fossil-fuel based energy generation. For energy storage, the reviewed solutions include hydrogen for long-term storage, battery-based system for short-term storage and regulation, and thermal storage using boreholes to meet heating demand. Where a fully renewable system cannot provide the required reliability, diesel can serve as back up generation. Some barriers to widespread adoption are technological, such as the need for specialized planning tools and equipment resilient to harsh weather, and social and institutional, such as the need for governmental support, subsidies, and appropriate legal frameworks.

## 1. Introduction

The Arctic covers an area of 20 million km$^2$ and is home to about 4 million people [1], [2]. There is no single geographical definition of the Arctic, and its boundaries vary depending on the purpose of the analysis. Common definitions include the area above the Arctic Circle (66° 34' N), the northern limit of tree growth (the Arctic tree line), or average daily summer temperatures below 10° C [2]. This region consists of 1500 off-grid remote settlements that host 1.6 million people across Canada, the United States, Russia, Denmark, and Norway [3]. Not all arctic communities are off grid, for example, communities in mainland Norway are generally connected to the national electricity grid, whereas remote regions such as Svalbard are not. Diesel is the primary fuel that is used for heating, electricity, and transport in these off-grid communities. In some communities, the ratio of fuel consumption for heat is almost 5 times higher than fuel consumption for electricity [4].

Arctic warming is primarily driven by global greenhouse gas emissions and associated climate processes rather than by emissions originating solely within the region. The Arctic is warming almost twice as fast as the global average [5], while local factors can further intensify its effects. Black carbon emissions from fossil-based energy systems are of particular concern, as they darken surfaces when deposited. In the case of sea and ice covers, it enhances the absorption of solar radiation, thereby causing an increase in melting rates [6]. From a global energy balance perspective, the Arctic's white sea ice reflects solar energy into space, thus serving as a massive temperature regulator. These shrinking ice covers can cause a dramatic increase in global warming [7].

Accelerated climate change in the Arctic region has significant impacts on the development of indigenous communities residing in these regions [8]. These communities face several challenges, particularly rapid environmental degradation and high energy costs. These challenges are closely connected to their heavy reliance on imported diesel, which exposes them to fuel-price fluctuations and high transportation costs while at the same time contributing to greenhouse gas emissions and local air pollution [9], [10]. Integrating renewable energy sources can help reduce these effects. Beyond environmental benefits, renewable energy generation in these communities offers greater energy independence, improved reliability, and long-term cost savings [11].

Renewable energy sources capable of supplying local autonomous power include wind, solar, tidal, geothermal and biomass energy. Other energy alternatives include cleaner forms of conventional fuels such as liquified natural gas and low-power nuclear power plants [12]. However, renewable energy systems face significant challenges in the Arctic regions. The remoteness of these locations, combined with technical difficulties in integrating time variable renewable sources in the system, drives up the cost of these systems significantly compared to the systems installed in more accessible areas [13]. In most cases, 35% of the capital cost is accounted for by delivery logistics and site installations [14]. Many Arctic communities are not connected by roads, which complicates the transportation, installation, and maintenance of energy infrastructure. Their remoteness and distance from existing transmission networks also make connection to national grids difficult, and in many cases, economically impractical. Without the support from the main grid, most of these communities use an islanded electricity microgrid configuration [15].

Low temperatures in these regions can significantly affect material properties, particularly below -40 °C. Such low temperatures can cause deformation of composite materials due to differences in thermal expansion rates and increase the brittleness of certain metals. In the case of wind turbines, icing on the blades can deform their shape, thereby reducing efficiency [15]. Installing infrastructure such as overland power lines and building foundation is more difficult due to the hazards associated with construction on frozen ground [16]. Permafrost, which consists of a permanently frozen ground layer beneath a seasonally thawing active layer, presents an additional challenge for energy infrastructure. As the active layer deepens, ground stability can decrease, potentially causing damage to building foundations, transmission poles and other infrastructure. Climate change is accelerating permafrost thaw, increasing the thickness of the active layer. In Svalbard, the active layer typically ranges between 1 and 2 m, although projections indicate that it may deepen, potentially exceeding 5 m in some locations under continued warming scenarios [17]. Addressing these challenges require policies, planning frameworks, and equipment specifically adapted to Arctic conditions, rather than relying on approaches developed for energy systems connected to large centralized grids [13].

This review examines the challenges and opportunities associated with integrating renewable energy systems in Arctic communities, with particular emphasis on remote and off-grid settlements. It reviews the unique environmental, technical and social constraints that influence energy system design and discusses strategies and technologies that can improve the reliability, sustainability and resilience of Arctic energy systems.

The review is organised as follows. Section 2 sets out the review methodology, from article identification and screening through to the profile of the included literature. Section 3 introduces the regions covered. Section 4 forms the main body of the review, examining conceptual energy systems for the Arctic: renewable generation and its cold-climate limitations, diesel and geothermal supply, battery and hydrogen storage, thermal systems and building-side demand, and the enabling technologies and policy frameworks that shape deployment in remote communities. Section 5 reviews the modelling tools used to design and optimise hybrid Arctic energy systems, comparing EnergyPLAN, MARKAL/TIMES, TRNSYS, HOMER, and MATLAB-based approaches. Section 6 concludes and identifies remaining research gaps.

## 2. Methodology

The methodology used here is based on the review by Li et al. [18]. This systematic analysis comprises three main stages: identification, screening, and eligibility. An overview of each stage is shown in Figure 1.

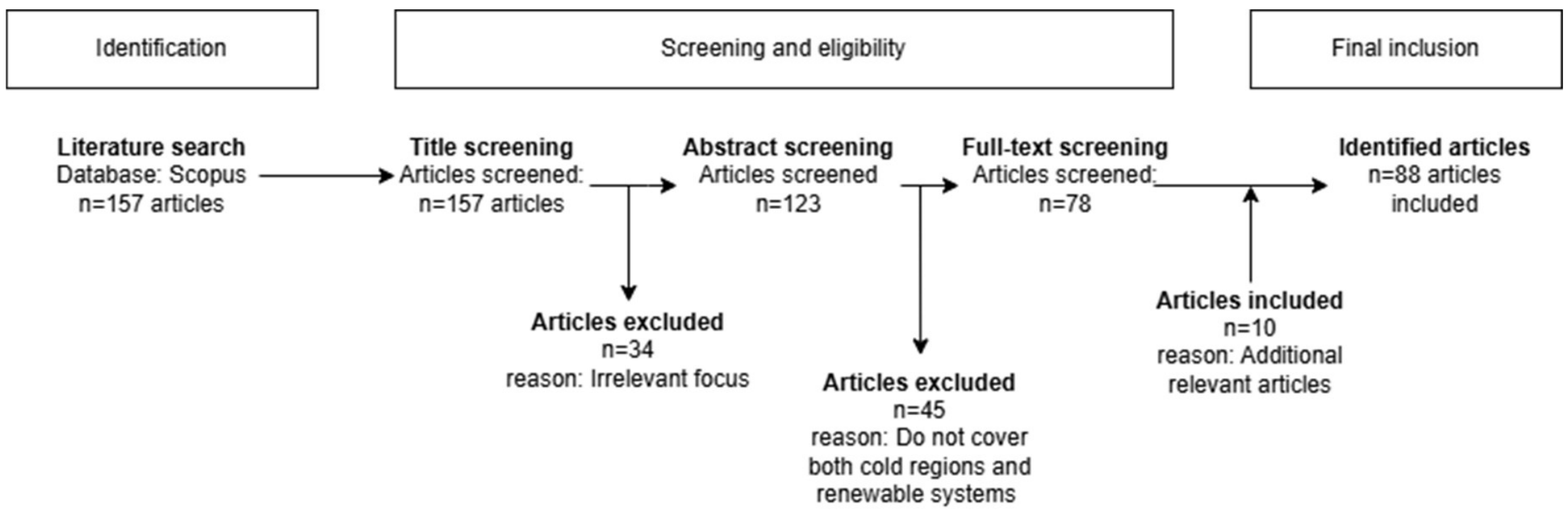

*Figure 1: Methodology on stages and screening steps used to identify relevant articles. Adapted from [18]*

## 2.1. Identification of Articles

The literature search was conducted on the Scopus database on the 22nd of October 2025, and the search query was limited to the article title, abstract, and keywords. The search query used is shown below:

```
TITLE-ABS-KEY ( ( "renewable energ*" OR "renewable energy source*" OR "sustainable
energ*" OR "clean energ*" ) AND ( "energy system*" OR "energy-system
model" OR "energy system modelling" OR "electric*" OR "decarboni*" ) AND
( "Arctic" OR "polar region" OR "cold climate" OR "circumpolar" ) ) AND PUBYEAR > 2015
```

Keywords used at this stage were limited to renewable, sustainable, and clean energy sources, as well as energy systems and modelling. The search was limited geographically using keywords such as arctic, polar, and cold-climate regions. The analysis was confined to articles published from 2015 to date. This search query returned 157 articles, which were further processed.

## 2.2. Screening, Eligibility, and Final Inclusion

The articles retrieved from the Scopus search query underwent a three-step screening process: title, abstract, and full-text screening. As the target of this review is energy systems in cold-climate regions, topics outside their scope were filtered out. In the first stage, a title screening was conducted on 157 articles, and those outside the scope of this review were removed. This process removed 34 articles, reducing the total to 123. In the next stage, an abstract screening was performed, and research articles that did not focus on cold-climate and energy systems were removed. This screening process removed 45 articles. The final screening included 78 articles that underwent full-text review. In addition, 10 relevant articles were added, bringing the total to 88.

## 2.3. Profile of the Reviewed Literature

This section provides an overview of the articles included in the review. Figure 2 shows the number of articles published each year and the number of studies gradually increased from 2015 onwards. Besides 2023, which was found to be an anomaly in which the Scopus search string yielded only three articles, a gradual increase in publications was observed, indicating growing research interest in these topics.

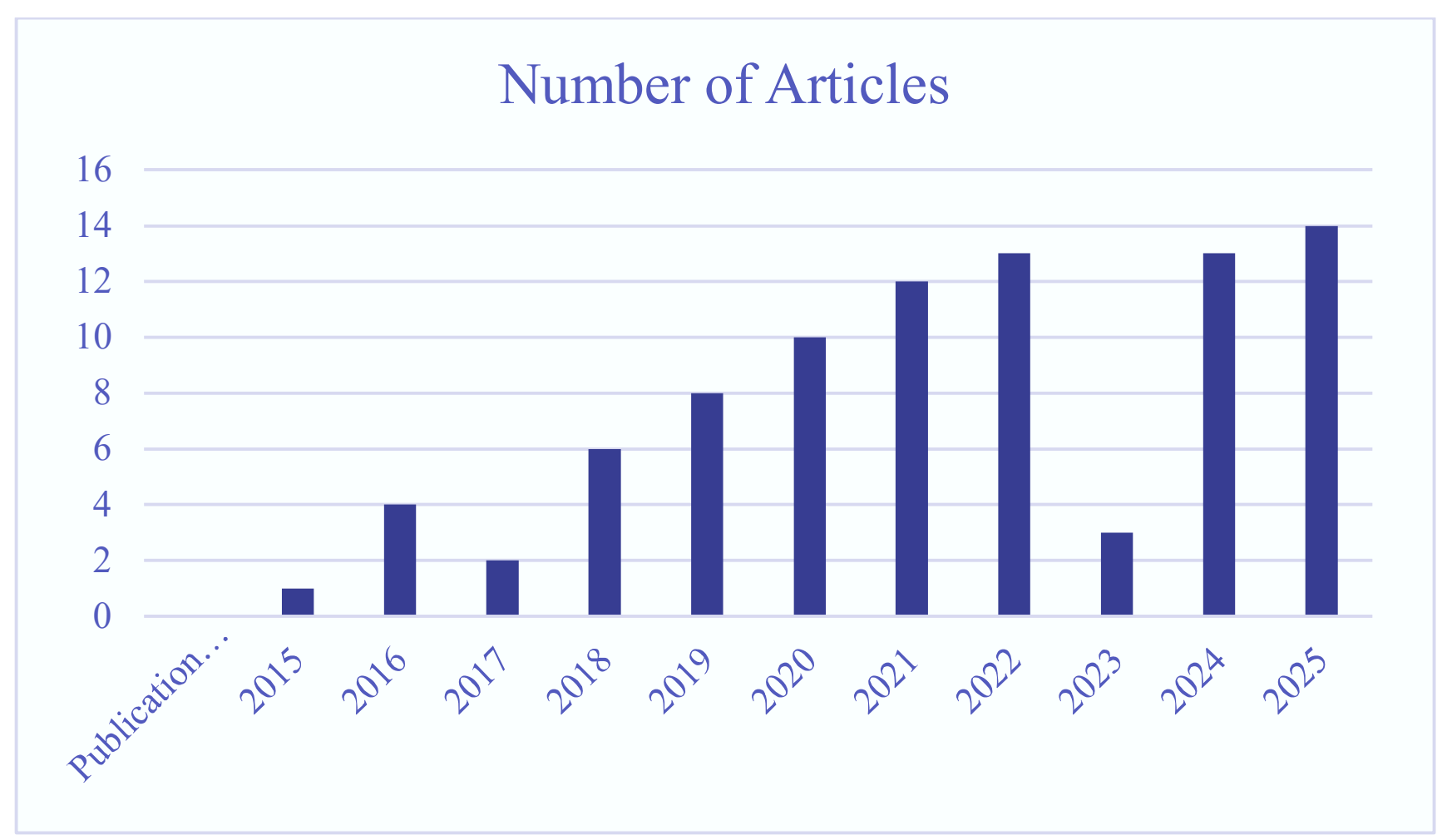

*Figure 2: Number of articles included from each year between 2015 and 2025.*

A geographical analysis of the articles included in this review shows that they cover a wide range of regions, with the majority from cold climates and Arctic regions. Figure 3 depicts the areas and the number of articles that address them.

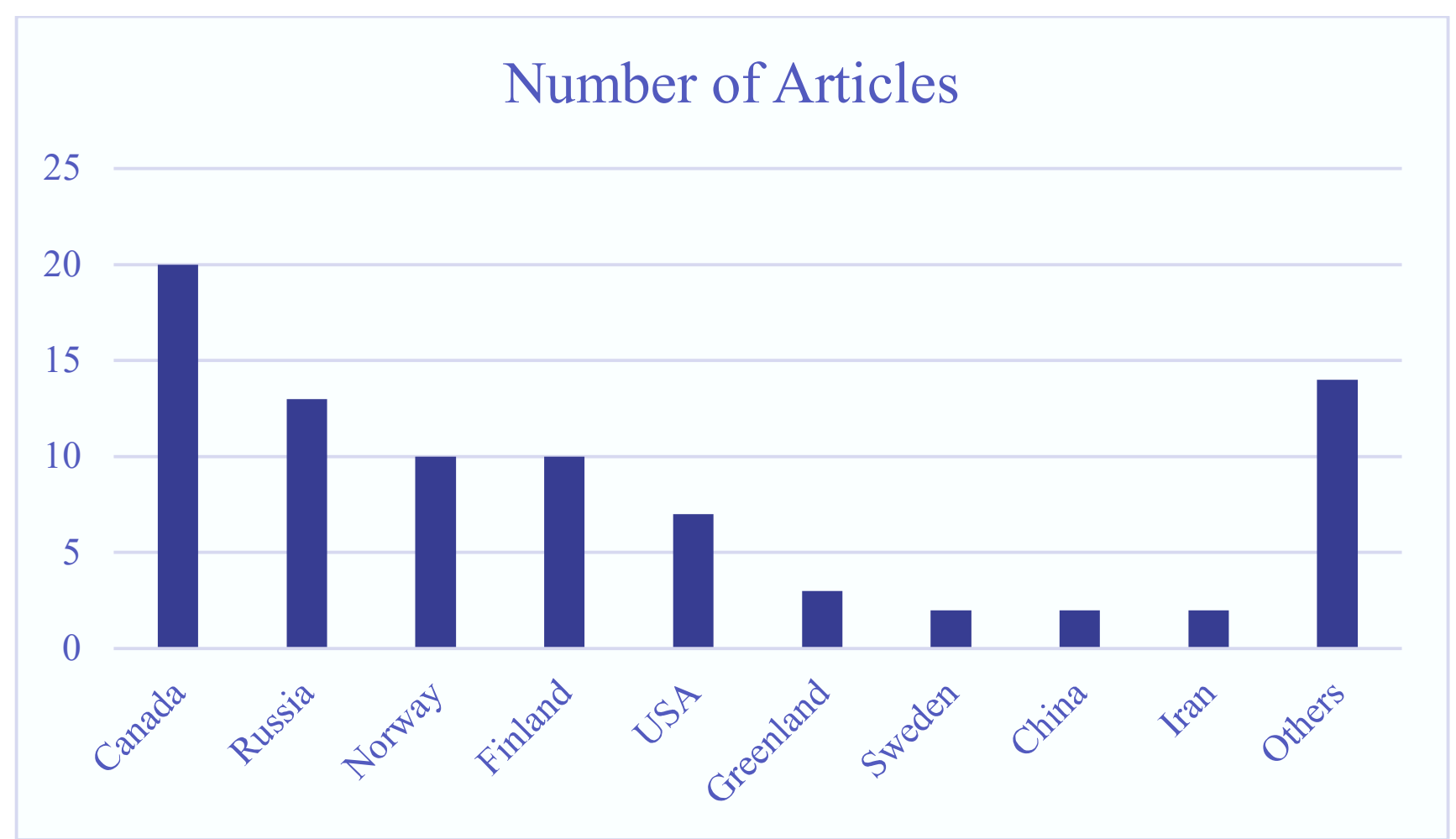

*Figure 3: Geographical distribution of articles included in the review.*

# 3. Overview of the Regions Covered

This section provides an overview of the locations covered in this review, their existing energy systems, and the challenges they face in adopting and integrating renewable energy technologies. The discussion focuses primarily on the Arctic and northern regions of Canada, Russia, the United States and Scandinavia. Difference in grid connectivity, fossil fuel dependence, infrastructure, governance and economic conditions create distinct challenges and opportunities for renewable energy development in each region.

### Norway

Renewable energy plays a central role in the Norwegian energy mix. Mainland Norway has an interconnected national grid and therefore benefits from the country's extensive renewable electricity generation, primarily from hydropower. Whereas Svalbard, Jan Mayen, Hopen, Bjørnøya and Ny-Ålesund are not connected to it and therefore depends primarily on diesel-generated electricity, while renewable alternatives are being developed.

Norway is a global leader in renewable energy deployment, with hydropower providing 99% of the electricity [8]. The main off-grid community in Norway is in Longyearbyen, Svalbard. It was founded in 1905 as a coal-mining community, and the power plant was recently upgraded from coal-fired to a diesel genset. The energy requirements for a 2100-year-round resident are about 40 GWh of electricity and 70 GWh of heat. Most electricity is consumed by the industrial sector, while household and the services sector consume most of the heating [19]. Several renewable energy installations are already operating in Svalbard as part of its transition towards a more diversified energy system. Svalbard Airport has a 137 kWp solar PV system and a liquid biogas (LBG) plant that started operating from 12 July 2026. Isfjord Radio Hotel operates a 300 kWp solar PV system integrated with a 405 kWh battery-storage system [17], [20].

## Canada

Canada has nearly 300 remote communities with a population of more than 225,000. These include all communities of Nunavut and some communities of north-west NWT and Yukon territories [6]. Approximately 75% of these indigenous communities rely heavily on diesel as their primary energy source, because diesel is considered a low-cost alternative to grid access [11], [21], [22]. However, diesel generators have high operational costs and a significant environmental footprint [6], [22]. On the other hand, the Canadian Arctic supports remote mining and industrial sites. There are 148 microgrids in Canada, of which 86% are in indigenous communities, 73% lack year-round road access, and 61% are fly-in communities [23].

## United States of America (Alaska)

According to the US Energy Information Agency, electricity in Alaska is generated from 47% natural gas, 39% hydroelectric, 10% petroleum-fired, and 4% non-hydroelectric renewables [24]. More than 200 communities are operating in a remote islanded grid configuration in Alaska, mainly because they are not connected to each other or to the north American electricity grid [25]. These communities vary in size, ranging from a few dozen to a few thousand people. The communities rely heavily on imported diesel to meet their energy demands [24], [26]. Usually, fuel costs, including transportation, make up nearly half of energy generation expenses [27]. A combination of imported fuel, difficulties in fuel delivery, and minimal economies of scale has resulted in energy costs and energy burdens that are significantly higher in remote Alaska than the US national average [28]. In Alaska, hydropower continues to play a significant role in the region's energy mix; however, without support from the statewide grid, its benefits remain largely localized. These hydropower plants were initially constructed to support mining activities in the region [24]. The non-hydroelectric renewables sector of Alaska’s energy portfolio includes more than 60 MW of wind capacity and 24 smaller community-based systems ranging from 100kW to 9MW. Alaska promotes diverse innovative energy solutions, including low-temperature geothermal, heat pumps, biomass combined heat and power, hydrokinetics, and fish oil as boiler fuel [27].

## Russia

The Russian territories in the Arctic begin with Franz Josef Land in the northwest and extend to Wrangel and Herald Islands in the east [4]. The Russian Arctic zone covers approximately 9 million $km^2$, which is one-fifth of the Russian Federation's territory [29]. The current population of the Russian Arctic zone is approximately 2,5 million people [12]. The region is considered one of the most polluted in the world [8], while accounting for up to 15% of the country’s GDP and a quarter of exports [29]. Compared with northern Norway, which is sparsely populated and connected to the national grid, and largely supplied by renewable sources, the Russian energy system is complex. First, the Russian Federation's Arctic zone is much larger than Norway's. Second, the distances between scarcely populated regions are much longer in Russia [8]. Currently, approximately 900 diesel power plants operate in the region, producing about 3 billion kWh annually [30]. Although renewable energy production is an important part of Russia’s energy policy for the Arctic regions, its deployment has so far remained limited [13].

# 4. Conceptual Energy Systems for Arctic Regions

This section provides an overview of the energy sources and system configurations examined in studies of Arctic regions, with particular attention to prominent renewable energy options such as solar and wind. Figure 4 presents a general outline of the energy systems explored in the reviewed literature.

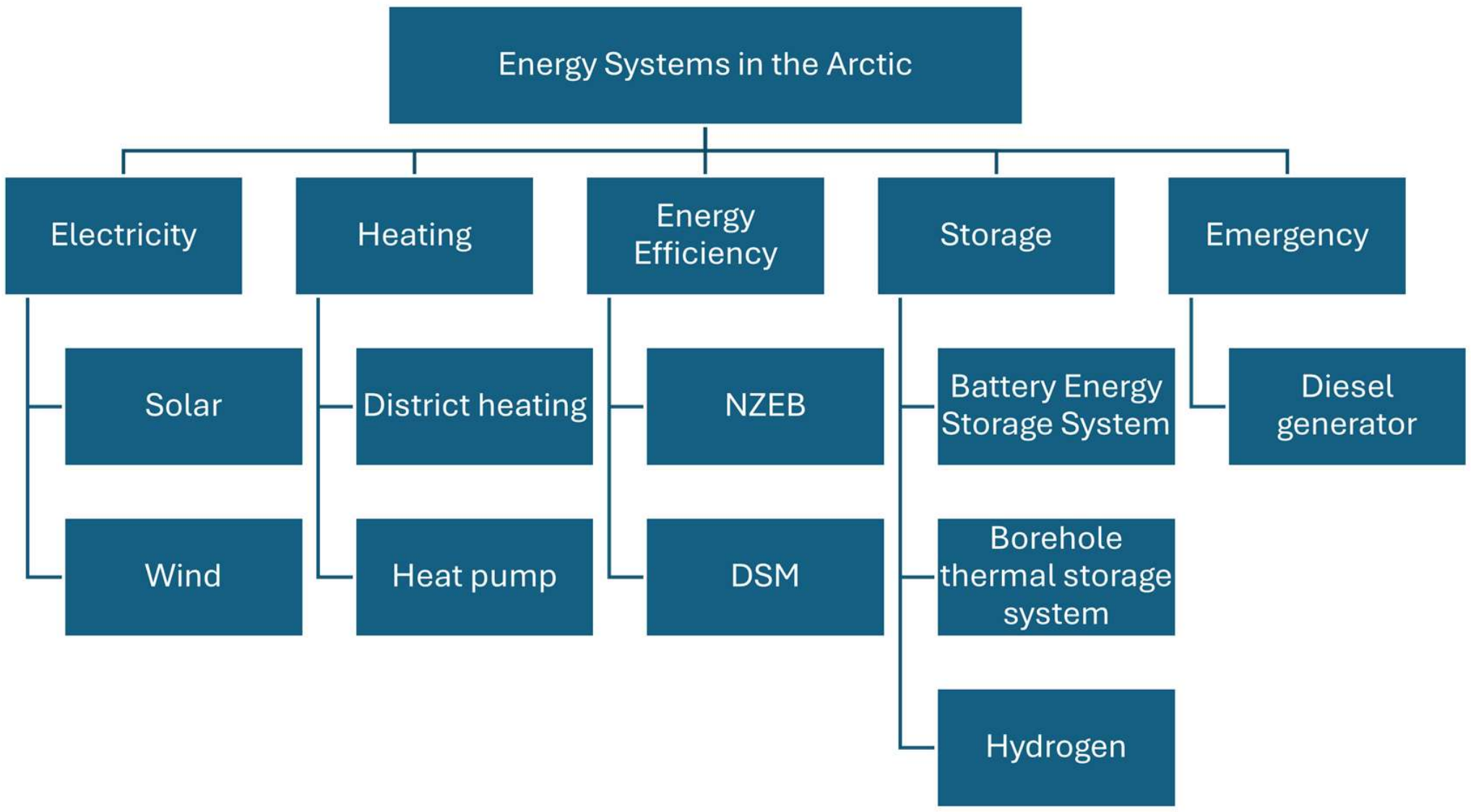

*Figure 4: A general overview of future energy systems in Arctic regions.*

Figure 4 summarizes the general direction proposed in the published literature for future Arctic energy systems. It does not represent a single existing system but rather an ensemble of complementary technologies drawn from different studies. Solar and wind are commonly proposed as the primary renewable energy sources, supported by battery, hydrogen and borehole thermal storage to address their variability and seasonal energy imbalances. Energy efficiency measures, including NZEB and demand side management, reduce overall demand, while district heating and heat pumps meet heating requirements. Diesel generators primarily act as emergency back up to maintain reliability during prolonged energy shortages.

In this review, electricity and heating demands are considered separately because heating constitutes a large portion of the total energy demand in the arctic conditions. Meeting this demand entirely through stored electricity would increase electrical storge requirements and introduce additional conversion losses. Li-ion batteries therefore provide short term electrical storage and fast response, hydrogen supports longer term energy storage, borehole thermal storage supplies heating demands and diesel serves as emergency backup. These technologies represent functions proposed across the reviewed studies rather than components of one existing Arctic energy system.

## 4.1. Renewable Energy Sources in the Arctic Energy Supply

Among the renewable energy resources available in the Arctic, solar and wind energy have emerged as most widely adapted for electricity generation in remote and off-grid communities. This section reviews the main characteristics of solar and wind energy in the arctic region, highlighting their potential and operational performance under cold climate conditions. It further explores the complimentary nature of these resources and key technical challenges associated with the integration of these resources into arctic conditions. It also reviews diesel and geothermal generation, along with battery and hydrogen storage systems used to manage variable renewable output. It then discusses thermal energy systems and building heating demand in cold climates. Finally, it presents the enabling technologies and pathways for the Arctic energy transition, including islanded microgrids, demand side management, nuclear energy, and the supporting subsidies and legal frameworks.

## 4.2. Solar Energy

Despite the harsh Arctic conditions, solar power is increasingly being harnessed across the region, although it currently accounts for less than 1% of total electricity generation. According to Dewitt et al. [1], approximately 30% of photovoltaic PV plants in the Arctic have capacities below 10 kW. The remaining 70% have installed capacities of 10-30 kW, with only a few acceding 100 kW. The annual-average daily solar irradiation across Arctic regions ranges from 2 to 5 kWh/m$^2$/day [8]. However, solar availability varies considerably throughout the year, with abundant sunlight during summer and little or no sunlight during the polar night period. Consequently, a PV system designed to meet the same annual electricity demand may require approximately three to four times as many panels as a comparable system in a southern location [33]. Wind generation may complement solar power during the dark winter months, although this relationship depends on local wind conditions. In this regard, solar PV systems require a substantial upfront cost compared to conventional diesel-based systems, but they have a very low operating costs [26]. The cost of solar PV has decreased drastically in recent years, so much so that price forecasts have systematically failed, as the decline in prices has been much faster than predicted [6], [34].

To maximize solar energy harvesting at low solar elevation angles, such as those in the Arctic, solar PV panels can be mounted on building walls rather than roofs [1]. These panels are sometimes installed in multidirectional arrays, which reduces the capacity factor but provides a more usable power output [24]. In these regions, east-facing panels maximize energy production in the morning, while west-facing panels maximize it in the evening. Moreover, solar PV panels can also be installed on tracking mechanisms to increase energy production [19].

Emerging PV technologies are focusing on improving efficiency, reducing costs, and enabling coupling with other energy generation or storage systems. These include bifacial solar panels, building-integrated solar panels, organic photovoltaics, and concentrated photovoltaics. For applications in the Arctic, bifacial photovoltaic panels offer an effective solution. Their ability to capture solar energy from both the front and rear surfaces significantly boosts their energy yield, especially on high-albedo surfaces such as snow-covered areas. For example, fresh and aged snow have albedo values of 0.6-0.9. Bifacial panels can take advantage of this reflected radiation by harvesting energy on the rear surface, thereby achieving a bifacial gain of 25-45% [35].

### 4.2.1. Degradation of Photovoltaic Panels in Cold Climates

According to Tonita et al. [36], one of the main barriers to the large-scale adoption of solar PV in cold-climate regions is the impact of temperature on long-term performance. Solar PV systems operating in cold climates are exposed to moderate to high snowfall, low temperatures, frequent freeze-thaw cycles, and strong seasonal variation in solar irradiance. Snow and ice accumulation on the panels exerts a non-uniform, irregular load, which can lead to cell cracking, while moisture ingress from melting snow and ice can lead to delamination and corrosion. Additionally, snow removal techniques, if used, can damage the module surface. Exposure to very low temperatures, such as -40 °C, can permanently reduce fracture strength, and cell connectors have been found to be susceptible to failure at these temperatures. Despite this potential structural weakening, cold operating temperatures increase the efficiency of solar photovoltaic panels. Tonita et al. [36] compared the degradation rates of PV panels in cold-climate regions with those in warm and temperate climates and reported that the lowest degradation rates occur in cold and polar climates [36].

Tonita et al. [36] conducted an extensive analysis of PV panel performance in cold-climate regions. They reported that annual bifacial gains for the evaluated system ranged from 27-34%, while double-axis tracking increased annual energy yield by 41%. They observed the performance of three double-axis tracking systems in Alaska and found that double-axis tracking increased energy production by 48-86%. The authors found that only one tracking system failed after 12 years of operation, further demonstrating the viability of bifacial and tracking systems in the Arctic's cold-climate regions. Damage to the panels were caused by moisture ingress at several sites. One panel was reported to have been destroyed by snow and ice, and by repeated seasonal freeze-thaw cycles [36]. In her doctoral thesis, Tonita reported that, to minimize reliability issues associated with moving mechanical components, tracker operation is typically disabled from October through March. Tonita also found that the double-axis tracking arrays exhibited a relatively low average degradation rate of approximately 0.4% per year over a 16-year observation period [37].

## 4.3. Wind Energy

Wind energy resources are widely available in the Arctic, especially along coastal regions. Unlike solar energy, wind is available year-round, but energy storage is required to cover generation deficits [23]. Cold regions can enhance wind turbine performance because air becomes denser at lower temperatures, allowing turbine blades to extract greater energy from the wind [1]. Recent studies have quantified the effects of temperature on wind power, showing that wind turbine performance can be 5% higher in winter and 10% lower in summer [38]. Wind energy projects in Arctic regions are expected to cost two to three times as much as comparable projects in temperate regions. While the typical lifespan of a wind turbine is around 20 years, it can be extended if operation and maintenance costs remain within acceptable limits [1].

### 4.3.1. Limitations on Wind Energy Scalability

Wind generation technologies provide a viable solution for decarbonizing electricity and heat sectors, but their deployment can be constrained by turbine scale. For example, commercially available wind turbines designed for cold regions are typically rated at 20-100kW or higher, whereas remote communities often have small populations and correspondingly lower energy requirements [9]. In their study, Paulin-Besset et al. [39] recommended Eocycle EOX S-16 wind turbines as suitable for cold-climate applications. These wind turbines use tilt-up tower technology with a crane-less hydraulic tower, enabling lower installation costs. These have a rated power output of 20-30 kW, a cut-in speed of 2.75 m/s, and a lifespan of 30 years. Their specified operating temperature range is -20 °C to 40°C, while the generator is housed within a fully enclosed, weatherproof enclosure under normal operating conditions (Eocycle, 2025. EOX S-16 Datasheet).

### 4.3.2. Blade Icing on Wind Turbine in Cold Climates

In cold-weather regions, one of the most significant challenges for wind turbines is the accumulation of snow and ice on their blades. Icing can alter blade shape and consequently the aerodynamic performance of wind turbines, thereby reducing their efficiency. Studies have shown that icing can cause approximately a 17-30% reduction in turbine performance [30].

Two types of ice commonly form on turbine blades: glaze ice and rime ice. Glaze ice is a hard, transparent layer that forms when liquid water freezes on cold surface. Rime ice is a rough, opaque deposit that forms when super-cooled water droplets freeze instantly upon contact with a surface, typically under cold, foggy, and windy conditions. Icing protection systems are generally divided into two categories: active and passive. Active systems require additional power (anti-icing systems are installed inside or outside the blades). Passive systems rely on coating the wind turbine blades with specialized material that reduce ice adhesion on blades, or applying black paint to the blades so that they absorb more solar radiation, thereby creating a natural heating effect [1].

## 4.4. Complementarity of Solar and Wind Resources

The two main sources of renewable energy, wind and solar, are complementary in their availability, as they exhibit variable generation patterns. This effect is further amplified in high-latitude regions, where solar energy is not available during the polar night (October to March). Inversely, the wind resource is at its strongest from September to April, but weaker during the summer months [19]. Multiple studies highlighted this complementary nature. Figure 5 is adapted from Hosseini et al. [40] in which the authors highlighted this effect.

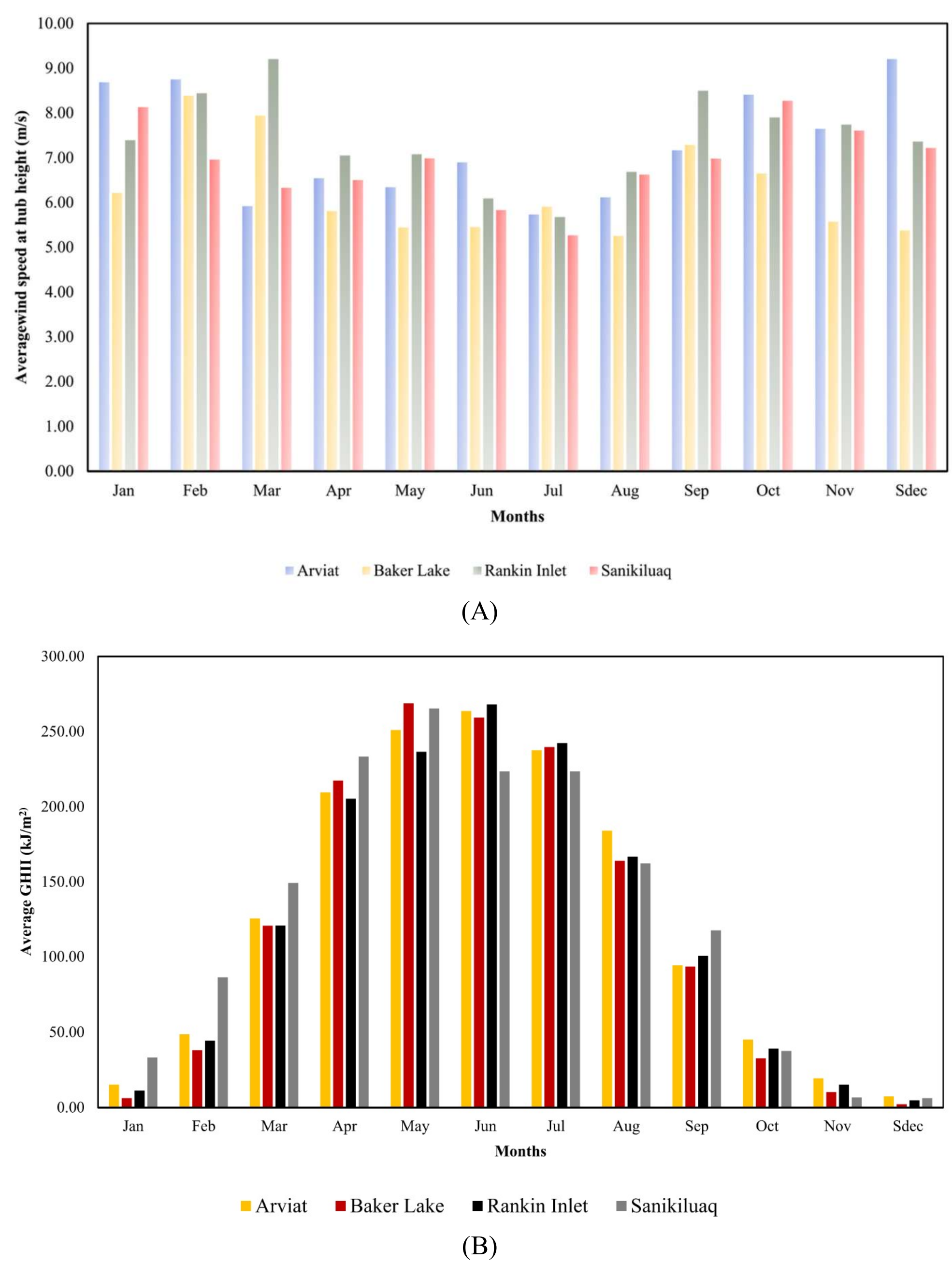

*Figure 5: Complementary nature of Wind and Solar resources. Adapted from [40].*

## 4.5. Challenges to Renewable Energy Deployment

Increasing the share of renewable energy in the energy mix is a fundamental step towards a low-carbon society. This increase in investment depends not only on the amount of new production capacity but also on the geographical location [25], [41]. The remoteness of an area increases the technical challenges of implementing renewable energy production units [13], [25]. Installations in remote regions such as the Arctic cost 3 to 4 times as much as in comparable installations in temperate areas. Beyond cost and logistics, a major technical challenge in deploying renewable energy sources is managing their variability. Energy storage can help maximize system efficiency and stabilize the grid; however, this increases the cost of energy supply [13]. Because of the variability of renewable energy sources, hybrid renewable energy systems require higher installed capacity to account for fluctuations in their availability [1]. For a successful transition, renewable resources should be utilized in a manner that is both technically feasible and economically viable [10].

## 4.6. Diesel Based Power Generation

Diesel is the primary energy source for almost 80% of the remote communities in the Arctic region [1], [15], [16], [31]. Diesel generators have proven their reliability in producing electricity under cold climate conditions, maintaining stable frequency and voltage. Only about 30% of the energy in diesel is converted to usable electricity, with the rest converted to heat, creating opportunities for cogeneration [32]. Most isolated renewable energy systems use diesel generators as emergency backup units to help meet peak demand during periods of high consumption and low production. On average, this backup capacity is set at 50% of the installed capacity [15].

Electricity generation from diesel is expensive due to volatile fuel prices and the need to transport fuel over long distances. The transport and storage of such fuels often result in oil spills and leaks in these vulnerable regions, further degrading the environment [1], [13], [31]. In two papers, Dewitt et al. proposed that it is advisable to transition to renewable energy technologies rather than continuing investing in diesel technology [10], [15]. One of the shortcomings of diesel generators is that they must operate at 20-30% of their rated capacity to operate properly. With high renewable energy penetration, the load on these generators is reduced, which can reduce grid stability and increase maintenance costs. However, operating these generators at higher loads limits renewable energy penetration, reduces fuel savings, and lowers overall efficiency [27].

To date, no Arctic facility has managed to operate through the winter without a diesel engine. A facility in Russia aims to achieve this by combining wind and solar power plants with lithium-ion and hydrogen energy storage systems [4].

## 4.7. Geothermal Energy

Geothermal energy, which remains relatively underexplored and underutilized compared with other renewable energy resources, is available in the Arctic region primarily along the Ring of Fire near the Aleutian Islands [1]. Geothermal power plants can provide a stable baseload supply for both electricity and heating demand [15]. Several studies have identified geothermal energy as a potential alternative for electricity and heat production in Longyearbyen, Norway. In Longyearbyen, preliminary test drilling has shown promising results, with ground temperatures in and around the settlement exceeding those observed in mainland Norway [19]. Senger et al. [42] conducted an extensive survey of geothermal gradients on Spitsbergen, Svalbard. They found that the gradient varies from 24 °C/km in the west to 55 °C/km in the southeast, with an average of 33 °C/km. In their study, Macmillan et al. [9] proposed direct geothermal heating as an alternative to conventional heating oil.

## 4.8. Energy Storage Systems in Cold-Climate Regions

Remote communities in cold climate regions that are not connected to a grid require energy storage system capable of stabilizing the voltage and frequency of islanded microgrids [24]. At the shortest time scale, this function is provided by spinning reserve from a generator or by a flywheel. The intermediate (daily or weekly energy demands) and seasonal time scales are often managed by battery energy storage systems [1]. Of the many energy storage technologies, electrochemical batteries, hydrogen, and latent heat provide notable advantages and potential applications in cold climate regions [75]. The majority of the studies in this review that investigate hybrid renewable energy systems include a battery energy storage system as one form of energy storage component of the system [11].

This section provides an overview of the energy storage technologies considered, with particular emphasis on lithium-ion batteries and hydrogen-based energy storage, including their suitability, performance and limitations in cold-climate regions.

### 4.8.1. Battery Energy Storage in Cold Climates

Energy storage systems play a critical role in mitigating the intermittency of renewable energy sources. Such systems enable the storage of surplus energy during periods of high generation and low demand, which can later be discharged during peak demand, ensuring the stability and reliability of the whole system [75].

Electrochemical battery systems are fundamental components of modern energy systems. This technology offers high efficiency, fast response times, and grid stabilization capabilities. These systems are particularly valuable for mitigating renewable energy intermittency and ensuring grid stability [75]. However, one of the main shortcomings of electrochemical battery storage systems is their higher degradation rates compared to other storage technologies [71].

#### 4.8.2. Hydrogen for Long-Term Storage in Cold Climates

Hydrogen is a high-energy carrier with significant potential for long-term energy storage. It is the lightest and most abundant element on Earth, yet it is rarely found in its pure form [76]. It is an environmentally safe alternative fuel and energy storage medium that can be produced from water using electricity [77].

In nature, hydrogen predominantly occurs in compounded forms like water ($H_2O$), methane ($CH_4$), or other hydrocarbons. 1 kg of hydrogen has equivalent energy content as 2.1 kg of natural gas or 2.8 kg of gasoline, which shows it has a very high energy content in terms of weight but in terms of unit volume, it has a very low energy content. A Liter of hydrogen at atmospheric pressure contains 2.8 kcal, while in liquified form it contains 2030 kcal. To increase the energy content of hydrogen, it is liquified; however, this process requires large amounts of energy (cooling to -253 °C) or high-pressure compression (300-500 atm) [76].

In the Arctic, the utilization of hydrogen presents several advantages and is attracting attention as an alternative to traditional fossil-based energy sources [29]. With increasing pressure on global oil and gas supply chains, developed countries are investing in alternatives to fossil-based systems, while efforts to reduce carbon footprints are growing. In this case, hydrogen presents a key solution, a view reinforced by the European Union's target of producing 20 million tons of hydrogen by 2030 [76], [78].

#### 4.8.3. Hydrogen Production: Technologies, Trends, and Challenges

The basic raw materials used in hydrogen production are water, hydrocarbons, and other natural materials that contain hydrogen (coal may contain 2-6% hydrogen by weight) [76]. Electrolyzer splits water into hydrogen using electricity, which is then stored in dedicated tanks and paired with a fuel cell to create an efficient energy solution [74]. A fuel cell performs the opposite reaction; it converts hydrogen to electricity and produces water as a byproduct. Unlike other battery types that must be recharged regularly, a fuel cell can operate continuously as long as hydrogen is supplied. There are two basic types of fuel cells: SOFC (solid oxide fuel cell) and PEMFC (proton exchange membrane fuel cell). SOFCs are the most widely used in stationary energy systems because of their simple design, but they operate at high temperatures (700-1000 °C). On the other hand, PEMFC uses a solid proton-exchange membrane as its electrolyte. Because of its simple, sealed construction, it reduces corrosion and extends the fuel cell's lifespan [76]. The primary drawback of fuel cell technology is its high cost. Fuel cell manufacturers remain optimistic as fuel cell costs have dropped tenfold and output has increased 12-15 times over the past 15 years [76].

Currently, hydrogen is produced from fossil fuels (48 % natural gas, 30 % oil, and 18 % coal) and transported via pipelines or storage mediums [79]. Natural reforming of methane gas is usually preferred as its oxidation does not release any $CO_2$. Hydrogen is classified into four different classes, i.e. green, grey, blue, and turquoise, depending on the specific carbon and toxic footprints associated with its production. The cleanest is green hydrogen, produced by electrolyzing water using renewable energy sources. Gray hydrogen is produced from fossil fuels, generating a large carbon footprint; for example, 1 ton of hydrogen produces 10 tons of $CO_2$. Blue hydrogen uses the same process as grey hydrogen, but the $CO_2$ produced in the process is captured and stored. Turquoise hydrogen is produced by thermal decomposition of natural gas, and instead of $CO_2$, this process releases solid carbon soot. Green hydrogen and turquoise hydrogen are of particular interest because these processes emit less or no $CO_2$. Currently, turquoise hydrogen is more economical, as pyrolysis requires much less energy than electrolysis [80]. The energy content of hydrogen is lower than the energy required for its production, with efficiencies ranging between 20 and 80% [76].

In this review, several articles proposed different methods of hydrogen production tailored to specific scenarios. Dvoynikov et al. [80] have proposed a gas-to-chem solution for processing hydrogen from ammonia, methanol, and cyclohexane. Zvereva et al. [81] showed the possibility of producing green hydrogen from electricity generated by a wind farm, the authors in Satymov et al. [38] explored the potential for producing e-fuels from wind resources in northern Finland. Similarly, Svendsmark et al. [79] found that northern Norway has significant potential for large-scale hydrogen exports if its energy resources are fully utilized.

#### 4.8.4. Hydrogen Storage and Transport

Hydrogen storage systems are generally categorized into physical and chemical storage technologies. Physical storage methods include compressing hydrogen and storing it as a liquid. Compressing hydrogen gas requires specialized high-pressure tanks, while storing liquid hydrogen requires maintaining extremely low temperatures, which complicates both storage and transport. Chemical storage methods bind hydrogen with metals or other inorganic compounds; however, these techniques have low energy density (in the case of metal hydrides, 1.5-3wt.% [80]) and involve additional complexity and costs associated with releasing stored hydrogen [4]. Dvoynikov et al. [80], for example, proposed the use of solid hydrides for hydrogen storage and transportation. These hydrogen storage systems make hydrogen a favourable candidate for long-term storage and, more importantly, seasonal storage. This is particularly important in high latitude regions of the Arctic, where wind and solar resources vary significantly throughout the year [25].

The transport of hydrogen presents unique economic and technical challenges. Constructing dedicated trunk pipelines for hydrogen is economically impractical; however, blending 20% hydrogen with natural gas ($CH_4$) can be used in the existing pipeline infrastructure. Another promising approach is transporting hydrogen as ammonia ($NH_3$), which offers advantages in handling and storage. Currently, the global network of hydrogen pipelines exceeds 15.000 km, with the majority in the United States. The longest hydrogen pipeline in the world, connecting Antwerp, Belgium, to Normandy, France, is 400 km long and operates at 100 atm [76].

#### 4.8.5. Hydrogen in Hybrid Renewable Energy Systems

Hybrid renewable energy systems (systems with multiple energy sources) are emerging as viable solutions for storing electricity as green hydrogen and supplying power to remote off-grid communities. This approach is more appropriate for Arctic regions where the surplus energy from variable renewable sources is converted to hydrogen for later use. In remote microgrid applications, hydrogen is often considered more advantageous than lithium-ion batteries due to its ability to deliver stable power and enable greater utilization of renewable energy [23]. However, it should be noted that converting stored hydrogen to electricity comes with operational challenges. Unlike lithium-ion batteries, these devices do not ramp up or down instantaneously, and frequent operations can accelerate degradation [23]. Mousa et al. [70] proposed hydrogen storage for managing daily loads (6-8 hours) and Li-ion batteries for peak shaving and sudden fluctuation demands. Similarly, the round-trip efficiency of the hydrogen storage system in one particular study was 26.7% which is relatively less than a battery storage system (81%), which shows energy losses in the hydrogen storage system when compared to battery storage systems (Li-ion) [23]. The advantage of hydrogen over other storage technologies is that it offers long-term storage of up to a week or more, while most popular storage devices support only a few days [59].

Multiple studies have recommended hydrogen as an essential component of future energy systems in remote, cold-climate regions. Chade et al. [59] proposed a hybrid wind-hydrogen-diesel energy system, whereas a wind-hydrogen system, which no longer requires diesel, requires significantly larger-capacity units. Temiz et al. [35] proposed a solar-driven hydrogen energy system for a data centre. In several studies, wind and hydrogen were proposed as essential renewable energy sources for future systems. In studies by Kalantari et al. [60] and Janke et al. [23] proposed energy systems based on wind and hydrogen. In a similar study, Ringkjøb et al. [19] suggested that wind and hydrogen be essential components of energy systems in Longyearbyen, Norway.

#### 4.8.6. Summary of Reviewed Papers

Table 1 summarizes the key findings of the studies on storage and hydrogen technologies.

*Table 1: Key findings from studies on storage and hydrogen technologies.*

| Article | Summarized results | Year |
|---|---|---|
| Chade et al. [59] | Implementing a Wind-Hydrogen-Diesel hybrid system in the Arctic proved most cost-effective, achieving 92% renewables and 85% diesel reduction with a 3.68-year payback. In contrast, a fully renewable Wind-Hydrogen system was not viable due to high costs. | 2015 |
| Grigoriev et al. [82] | Hybrid power systems integrating solar and wind energy with methanol-based fuel cells provide a reliable, emission-free alternative to decommissioned radioisotope generators for remote Arctic infrastructure. | 2019 |
| Nazarova et al. [12] | This study demonstrates that renewable energy deployment in remote Russian Arctic communities is economically feasible, finding that wind-diesel hybrid systems are currently more financially viable than wind-hydrogen storage options. | 2019 |
| Shulga et al. [83] | Hydrogen fuel cells, low-power nuclear plants, and modular renewables offer the most efficient and eco-friendly energy solutions for the Russian Arctic. | 2020 |
| Dvoynikov et al. [80] | The study proposes Gas-to-Chem solutions for Arctic oil and gas development, using ammonia, methanol, and cyclohexane to enable low-carbon hydrogen processing. | 2021 |
| Kalantari et al. [14] | The study shows that combining fuel cells with battery storage offers the most cost-effective solution compared to using either technology alone. | 2021 |
| Obara and miyazaki [84] | The study solves electricity fluctuation and frequency stability issues by modelling a microgrid that uses high-efficiency fuel cell plants (SOFC-TCC) to balance large-scale photovoltaics. | 2021 |
| Temiz and Dincer [35] | The study successfully developed a unique, solar-driven hydrogen energy system for a data centre with an 11.2 MW electrical capacity in the Arctic Region, achieving 100% self-sufficiency. | 2022 |
| Kalantari and Ghoreishi-Madiseh [60] | Renewable energy system utilizing a hydrogen power fleet combined with hybridized battery/fuel cell storage provides the most economical solution. | 2022 |
| Galimova et al. [5] | Transitioning Greenland to an economically viable 100% renewable energy system leverages vast wind resources to produce hydrogen as an essential precursor for domestic e-fuels and high-value sustainable energy exports | 2024 |
| Sekretaryovyuri et al. [76] | Fuel cells could compete with mini-CHPs in Arctic decentralized systems only if costs drop below $100/kW, as they are currently twice as expensive. | 2024 |
| Svendsmark et al. [79] | If Northern Norway manages to exploit its energy resources, there is significant potential for large-scale hydrogen exports, and becomes profitable above 90 €/MWh. | 2024 |
| Zhdaneev et al. [4] | The findings confirm that hydrogen storage is critical for seasonal energy balancing in the Arctic. | 2024 |
| Zvereva et al. [81] | Producing green hydrogen from electricity generated by the Kola Wind Farm is technically feasible and could significantly reduce carbon emissions. | 2024 |
| Ju et al. [78] | The results showed that the GSHP with PV and waste heat from Hydrogen production has the lowest Life-Cycle Cost. | 2025 |
| Mckinley et al. [25] | Near-term capacity expansions enabling 50% renewable generation are found to reduce the total cost, while achieving 75% renewable generation is cost-competitive, but reaching 90% or higher remains uneconomical due primarily to high capital costs, which also makes seasonal long-duration hydrogen storage, requiring volumes up to 120,000 kg for complete seasonal shifting, impractical. | 2025 |
| Satymov et al. [38] | The article explores the potential to produce e-fuels from wind power in northern Finland. | 2025 |
| Janke et al. [23] | The presented model for hydrogen energy storage in northern microgrids demonstrated improved accuracy compared to standard tools, favouring wind integration over solar due to feasibility constraints. | 2026 |

## 4.9. Thermal Energy Systems in Cold Climates

Thermal energy systems play an important role in meeting the high heating demand of cold-climate regions. This section reviews approaches for thermal energy storage and heat supply, as well as strategies for reducing and managing thermal demand through heat pumps, district heating, building efficiency and net zero emission buildings.

### 4.9.1. Thermal Energy Storage

Thermal energy storage is a key component of energy systems in cold-climate regions and can offer a cheaper alternative to conventional electrochemical batteries [14]. When wind or solar energy is integrated with thermal storage, renewable generation capacity can be sized above the immediate energy demand. This additional capacity allows the system to meet the current load while charging the thermal storage during periods of high renewable generation, making stored heat available during periods of low generation [39]. Available heat output and thermal losses are important considerations in the design of thermal storage units. For large-scale applications, various types of thermal storage units, such as boreholes, rockpiles, and aquifers, can be used [14].

There are three primary methods of thermal energy storage: thermochemical storage using reversible chemical reactions, sensible heat storage, and latent heat storage [3], [4]. In sensible heat, the temperature of a heat carrier material increases or decreases. Heat carriers used in these systems could be liquid (water, mineral oil, salt solutions, liquid metals, and alloys) or solids (stone, concrete, sand, and bricks). In contrast, latent heat is stored in materials with phase change properties, where heat absorption results in a phase change in the storage material [4]. Polar Night Energy, a Finnish company, has developed an innovative mega-watt class solution for Arctic regions using soapstone powder, which has been in operation since 2022. The material is stored in thermally insulated containers at temperatures of up to 1000 °C [4].

### 4.9.2. Borehole Thermal Energy Storage

A promising thermal energy system for cold climate regions is the borehole thermal energy storage BTES. These systems utilize the subsurface as a heat storage medium through an array of borehole heat exchangers BHEs [85]. BTES is considered more appealing than other seasonal storage options because it combines affordable, large-scale, and long-term operation with the flexibility to expand capacity by drilling additional boreholes [11], [52]. Several studies have shown how a distributed network of thermal energy storage at point-of-use sites in the residential sector can assist in deep decarbonization efforts. Some communities in Alaska have installed dispatchable thermal load in individual residences [24]. Storing thermal energy on-site at the point of use is one of the most cost-effective distributed energy storage options in cold climate regions [11]. Other studies have proposed medium-depth BTES systems with efficiencies exceeding 80% as compared to shallow depth systems [85].

### 4.9.3. Heat Pumps

Heat pump technology, when integrated with renewable energy sources, is considered an energy-efficient solution that offers significant environmental benefits [86], [87], [88]. These devices transfer heat from one temperature level to another, requiring far less electricity to produce heat compared to conventional heating devices [89]. These can reduce heating electricity consumption by more than 3 times [4].

Heat pumps for space heating are typically divided into residential units and utility-scale district heating systems [88]. In residential space heating, these devices play a dominant role, although their efficiency depends on ambient temperature and heating demands. On the other hand, industrial heat pumps usually use the waste heat of industrial processes with stable temperatures [88]. These devices help reduce $CO_2$ emissions in space heating by using renewable energy sources from the environment, such as air (aerothermal), water (hydrothermal), and ground (geothermal) [90].

There are two main types of heat pumps, i.e., air-sourced heat pump ASHP and ground-sourced heat pump GSHP. Both systems draw heat from either air (as in ASHP) or the ground (as in GSHP) and convert it into high-grade thermal heat using electricity [55]. ASHP is a highly efficient alternative for space heating, offering energy savings and reduced emissions. However, their performance is heavily dependent on outdoor air temperature, which limits their application in cold regions [86]. One of the main reasons for the widespread adoption of ASHP as an auxiliary system is that it eliminates pollution from older coal- and biomass-fired heating systems while avoiding high operating costs [91]. Unlike ASHPs, GSHPs are less affected by outdoor air temperature because they utilize a relatively stable thermal energy from the ground. [55]. This makes GSHP a more favourable heating option than ASHP in cold regions [78], [92].

The coefficient of performance COP characterizes the performance of heat pumps. It is the ratio of the heat generated to the electricity consumed, and it varies with the temperatures of the source and sink. An ideal heat pump would have a COP of 8.8 to 10.1; however, in real-world applications, it lies in the range of 3 to 5 [88]. In a study by Bogdanov et al. [88] the authors showed that accurate COP modelling improves heat pump planning and showed large variations in results with a constant COP.

Renewable technologies like solar PV can be easily integrated with heat pumps, thereby increasing the share of primary energy from renewable sources and improving heat pump performance indices. Solar-assisted heat pumps SAHP are such devices and are classified into direct expansion DX-SAHP and indirect expansion IDX-SAHP [55], [86]. By 2050, it is expected that heat pumps will contribute a substantial share of industrial heat demands, partially substituting the direct use of biomass fuels in industrial processes [88].

## 4.10. Thermal Loads and Building Efficiency in Cold Climates

This section of the review includes studies that address thermal demands and residential energy requirements in cold-climate regions. These two areas were grouped based on the overlap of the research questions addressed by these studies.

### 4.10.1. District Heating

District heating (DH) is recognized as an efficient and sustainable approach to providing space heating and domestic hot water (DHW) in cold regions and can support the transition toward a sustainable and decarbonized heating system [46], [85]. A district heating system is a network of insulated pipes that distributes heat to consumers connected to a grid through heat exchangers [89]. It is widely used for heating in northern Europe, including Finland, Norway, and Sweden [78], [93]. The residential sector accounts for approximately 11.7% of GHG emissions, with space heating and DHW accounting for nearly 80% of household energy use in cold regions [72], [78], [87].

To decarbonize the district heating sector, renewable energy sources must replace conventional heat sources. This transition necessitates the integration of variable renewable energy sources that require seasonal storage solutions [85]. Integrating solar energy with DH has been found to offer additional benefits, as such systems can provide the electricity required to operate the heat pump and use excess solar energy to increase the thermal storage temperature [52], [78], [94]. The storage of thermal energy becomes essential in configurations where, for example, a variable-energy source such as solar photovoltaic or solar thermal collectors is used. These devices can produce surplus energy in the summer months when heating needs are minimal, and this excess can be stored in short-term or seasonal storage [78].

Several studies have found that DH will play an increasingly important role in future 100% renewable energy systems by enabling flexible integration of electricity and heating through cogeneration of heat and power (CHP), heat pumps, and electric boilers. The current DH systems are 3rd-generation district heating (3GDH) systems that typically operate at temperatures above 80 °C. Such high temperatures cause grid losses usually exceeding 20%. The newer iteration of these systems, the 4th-generation district heating (4GDH), overcomes this shortcoming by operating at a lower forward temperature of 55-65 °C (return temperature of 25 °C). Sorknæs et al. [46] showed that going from 3GDH to 4GDH decreases the primary energy consumption by around 4.5% and the costs of the system by 2.7%. 4GDH systems are designed to facilitate better integration between the sectors, reduce grid losses, and allow for the integration of renewable energy sources [85]. Due to changes in forward and return temperatures in 4GDH, the production and heat storage technologies are affected compared to 3GDH systems [46]. The newer generation of district heating systems allows the low-grade renewables and waste heat to be utilized [78]. The coefficient of performance of heat pumps is also expected to be higher in 4GDH [46].

Currently, heating generated by electricity is approximately 72%, 32% and 26% in Norway, Finland, and Sweden, respectively (45% of building space is heated by DH in Finland [34] and 70% of hotels in Norway and Sweden are heated by DH [89]). In these regions, a medium-sized municipality can reduce its annual electricity consumption by 10% and peak electricity demands by 20% by converting from electric to non-electric heating [93]. In studies by Rehman et al., their first article [52] showed that solar district heating can achieve 65-90% of renewable fraction, and in the second study, they analysed technical failures faced by solar district heating systems [95].

### 4.10.2. Summary of Reviewed Papers

Table 2 summarizes the key findings from studies on space heating and thermal energy storage.

*Table 2: Key findings from studies on space heating and thermal energy storage.*

| Article | Summarized results | Year |
|---|---|---|
| Rehman et al. [52] | The results show that an optimized solar district heating system can achieve 65-90% renewable fraction, but technical failures can degrade performance. | 2017 |
| Rehman et al. [95] | This study analyses technical failures in Nordic solar district heating systems, finding that storage tank de-stratification has the most significant adverse impact, with other issues such as pump control changes and reduced heat pump COP also significantly increasing electricity demand. | 2018 |
| Sorknæs et al. [46] | Going from 3GDH to 4GDH reduces the primary energy consumption of the entire energy system by around 4.5% and the system costs by 2.7%. | 2020 |
| Pike and Kummert [54] | The study revealed that thermal energy demands (space heating and domestic hot water) account for approximately half of the community's building-related fossil fuel consumption and GHG emissions. | 2021 |
| Sambor et al. [26] | Adding 1.5 kW of solar PV with optimized Water Reuse System control can cut lifetime costs by over 13% and diesel microgrid bills by up to 37%, while using excess wind energy can reduce total project costs by 68%. | 2022 |
| Usman et al. [92] | This study shows that climate and tariffs strongly influence system design, with PV and ground-source heat pumps emerging as optimal solutions for cold regions. | 2022 |
| Bogdanov et al. [88] | Accurate COP modelling improves heat pump planning and cuts energy demands. | 2024 |
| Hyvonen et al. [61] | Solar PV systems can cost-effectively provide renewable electricity to data centres in cold climate regions. | 2024 |
| Shi et al. [96] | The paper proposes a data-driven distributionally robust optimization-based demand response model using appliance flexibility and thermal storage to reduce peak-to-valley loads and renewable curtailment under uncertainty. | 2024 |

### 4.10.3. Net-Zero-Emission Buildings NZEB

The International Energy Agency reported that the building operations account for approximately one-third of global electricity consumption and 26% of global $CO^2$ emissions. Building energy demand has increased at an average annual rate of approximately 1% over the past decade [96]. In the European Union, buildings are the largest energy consumers and GHG emitters, representing 40% of total energy consumption [49], [52]. In cold-climate regions like those of the Arctic and Northern America, more than 60% of energy use is attributed to space heating [87]. In developed economies, people spend around 87% of their time indoors, further emphasizing the importance of reducing energy consumption and emissions associated with buildings [97].

A net zero emission building NZEB is defined by the US Department of Energy as *"an energy-efficient building where, on a source energy basis, the actual annual delivered energy is less than or equal to the on-site renewable exported energy"*. Under this definition, an NZEB balances its annual energy consumption with energy generated from on-site renewable sources. Broader net-zero definitions may also consider the embodied energy and emissions associated with building construction [87]. These are low-energy buildings in which annual energy production balances yearly energy consumption through renewable energy sources [98]. The criteria used to evaluate NZEB performance vary according to the system boundaries, energy-balance methods, and objectives of the assessment [99][99]. These indicators are shown in Table 9 in the Appendix.

Renewable energy technologies such as solar, wind, and hydropower, together with energy carriers and storage technologies such as hydrogen, are important enablers for buildings to reduce or eliminate energy consumption and become NZEB [87]. Building energy requirements generally include electrical energy and thermal energy for space heating, domestic hot water, and, in some cases, cooling. These demands can be partly or fully supplied using renewable energy sources [94].

Different technologies can be used to meet or offset individual building energy demands and support the achievement of net zero status. Technologies used for space heating and domestic hot water include evacuated tube and flat solar collectors, concentrating solar collectors, heat pumps, and combined cooling heating and power systems [99]. Astudillo et al. [49] found that an air-source heat pump was the most cost-effective solution in cold regions. In another study by Rehman et al. [100] the authors show that if all demands are included, i.e., heating, cooling, and plug-in loads, then it is difficult to reach the PEB (positive energy building) levels.

The building envelope is particularly important for reducing space-heating demand. Energy consumption can be reduced by more than 30% through improved air tightness [49], [101]. Additional envelop measures, including improved insulation, high-insulation windows, reducing thermal bridging and heat recovery ventilation, can further decrease heat loss. In Arctic regions, these measures reduce both annual energy consumption and peak heating demand, allowing heating systems, renewable generation, and energy-storage units to be designed with smaller capacities. Lower heating demand also reduces dependence on diesel generation during periods of low renewable-energy availability, making net-zero targets more achievable. Both Asaee et al. [87] and Teamah et al. [101] identified building envelop retrofits as essential for improving energy efficiency. The residential sector can further support energy-management strategies through advanced metering infrastructure and demand-side management. These systems can shift flexible electrical and thermal loads to periods of higher renewable generation, thereby improving the reliability and stability of the wider energy system [102].

#### 4.10.4. Summary of Reviewed Papers

Table 3 summarizes the main findings of the studies on net-zero energy buildings reviewed in this section.

*Table 3: Summary of the main findings from studies on net-zero energy buildings.*

| Article | Summarized results | Year |
|---|---|---|
| Niemela et al. [98] | The study found that the NZEB targets can be achieved cost effectively in deeply renovated Finnish educational buildings by combining GSHP systems, PV panels, roof insulation, and efficient windows, without requiring wall insulations. | 2016 |
| Asaee et al. [87] | Achieving net-zero emissions in cold-climate housing requires more than electrification; it demands a combined approach of deep retrofits, renewable energy integration, and energy-supply decarbonization tailored to each country. | 2018 |
| Harkouss et al.[99] | For heating-dominant climates, the optimal NZEB configuration consists of biodiesel generators for electricity and steam, supplemented by solar thermal, PV, wind turbines, and ground-source heat pumps to achieve optimal load matching. | 2019 |
| Rehman et al. [100] | If heating, cooling, and plug-in loads are included, achieving a positive energy building PEB is challenging, with energy system investments comprising 47–62% of the life-cycle cost, and the remainder is composed of operational expenses. | 2022 |
| Teamah et al. [101] | Major building retrofits can cut energy use by 52% and $CO_2$ emissions by 82 tons, with a 5-year ROI, while adding renewables further lowers emissions but extends payback time. | 2022 |

| Kianpoor et al. [102] | The novel Home Energy Management System framework for EV charging in Northern Norway reduced total electricity costs by 15.4%, EV charging costs by 52.2%, and peak power consumption by 23.19% through smart load shifting to low-demand periods. | 2024 |
|---|---|---|

## 4.11. Enabling Technologies and Pathways for the Arctic Energy Transition

Building on the energy generation, storage, and thermal technologies discussed in the preceding sections, this section reviews system-level approaches, advanced technologies, and enabling measures that can support the transition towards low and zero emission energy systems in remote Arctic communities. The reviewed studies address islanded microgrids and advanced control, demand side management, innovative energy conversion and storage concepts, nuclear energy technologies, subsidy mechanisms, and transition frameworks. Solutions developed for temperate regions may not be directly transferable to Arctic conditions, where prolonged low temperatures, pronounces seasonal variation in renewable energy availability, and the operational constraints of isolated systems can limit renewable energy integration [70]. The following subsections examine how these technologies and measures can improve system reliability, reduce diesel consumption and support the Arctic energy transition.

### 4.11.1. Islanded Microgrids for Remote Arctic Regions

In the remote arctic regions, most communities lack access to a centralized electricity grid. Under such conditions, islanded microgrid provide a viable solution for local energy generation and distribution [103]. A microgrid is a network of distributed generators, energy storage units, and loads that operate as a single controlled entity [1], [10].

Arctic microgrids are increasingly incorporating hybrid renewable energy sources, such as wind and solar. As these sources are inherently variable in nature, microgrids can manage fluctuations in renewable generation through coordinated control of generation, storage and demand, thereby improving system resilience and reliability [1], [103]. Studies on advanced methods to improve microgrid stability, including those by Das et al. [6] and Hernandez et al. [103], showed the advantages of using variable-speed generators and grid-forming inverters in cold-climate regions, respectively. However, very high renewable penetration can introduce additional operational and economic challenges. Punyam et al. [73] showed that renewable penetration beyond 70% can lead to sharp increases in energy curtailment and overall system costs.

Other advanced technologies considered in this review include compressed air energy storage and ocean thermal energy conversion. Compressed air energy storage uses underground or underwater cavities to store energy. Hunt et al. [90] proposed an innovative isothermal deep-ocean compressed-air energy storage system. In a similar study, Temiz et al. [77] proposed ocean thermal energy conversion devices that are more effective in cold regions.

### 4.11.2. Demand Side Management DSM

Demand side management (DSM), also called demand side response (DSR), refers to strategies that optimize electricity consumption on the user side. The primary goal of DSM is to reduce peak demand, shift energy consumption to off-peak hours, and encourage the adoption of energy-efficient technologies. These strategies help balance supply and demand, lowering operational costs and minimizing the need for additional power generation infrastructure [25].

The two main components of DSM are demand-side-optimization (DSO) and demand response optimization (DRO). DSO is responsible for operating and maintaining the local electricity grid and ensuring reliable power delivery. DRO manages demand and supply by providing financial incentives. These two components form the core of DSM [26].

Sambor et al. in two of their studies [26], [104] showed that investing in 17 kW of solar PV for an Arctic container farm can cut annual costs by up to 18%, while in the subsequent study, they showed that adding 1.5kW of solar PV with optimized water reuse system control can cut lifetime costs by over 13% and diesel microgrid bills by up to 37%.

Another example of DSM implementation is through using an intelligent automatic control system (IACS). It consists of a diagnostic unit that provides data acquisition and supervisory control, a control unit that distributes energy in the system, a forecasting unit, and an ice prediction unit [30]. Similarly, Elistratov et al. [30] presented an advanced intelligent automatic control system that reduced fuel consumption by 22%.

### 4.11.3. Nuclear Energy Technologies for Remote Arctic Regions

Radioisotope thermoelectric generators have been extensively used since the 1960s to power lighthouses, navigation beacons, remote meteorological stations, and remote sensing installations that require an independent power source. Grigoriev et al. [39] show that, unlike uranium fuel used in nuclear reactors, strontium-based generators do not pose a nuclear hazard. They provide an analysis of 700 units deployed in the 1990s. The amount of heat generated by a single unit over its half-life is equivalent to the combustion of approximately 4 tons of diesel.

Small modular reactor SMR represents a new generation of nuclear technology, designed to operate at atmospheric pressure and incorporate passive safety controls. These types of reactors are explosion-resistant and require only light water for operation. However, the reliability of these passive safety features has not yet been verified, and thus their use has not been approved. These reactors have a low levelized cost of electricity LCOE which is comparable to a low-cost large-scale hydroelectric power plant and 40-90% lower than the LCOE of diesel generation [32]. These types of nuclear reactors are developing rapidly in Russia, with capacities ranging from 0.1 to 200 MW [83]. Zhdaneev et al. [4], reported that the village of Pevek in Russia has a floating nuclear plant in operation, equipped with 2 KLT-40 reactors providing 35 MW of electrical power and up to 85 MW of thermal power (see Figure 6).

Even if SMRs are approved for deployment, the key challenges include the risk of unsupervised nuclear material, high greenhouse gas GHG emissions during SMR manufacturing, and strong public scepticism towards nuclear power. In 2015, Nunavut residents in Canada opposed a uranium mine proposal; stronger opposition to nuclear reactors is anticipated. [32].

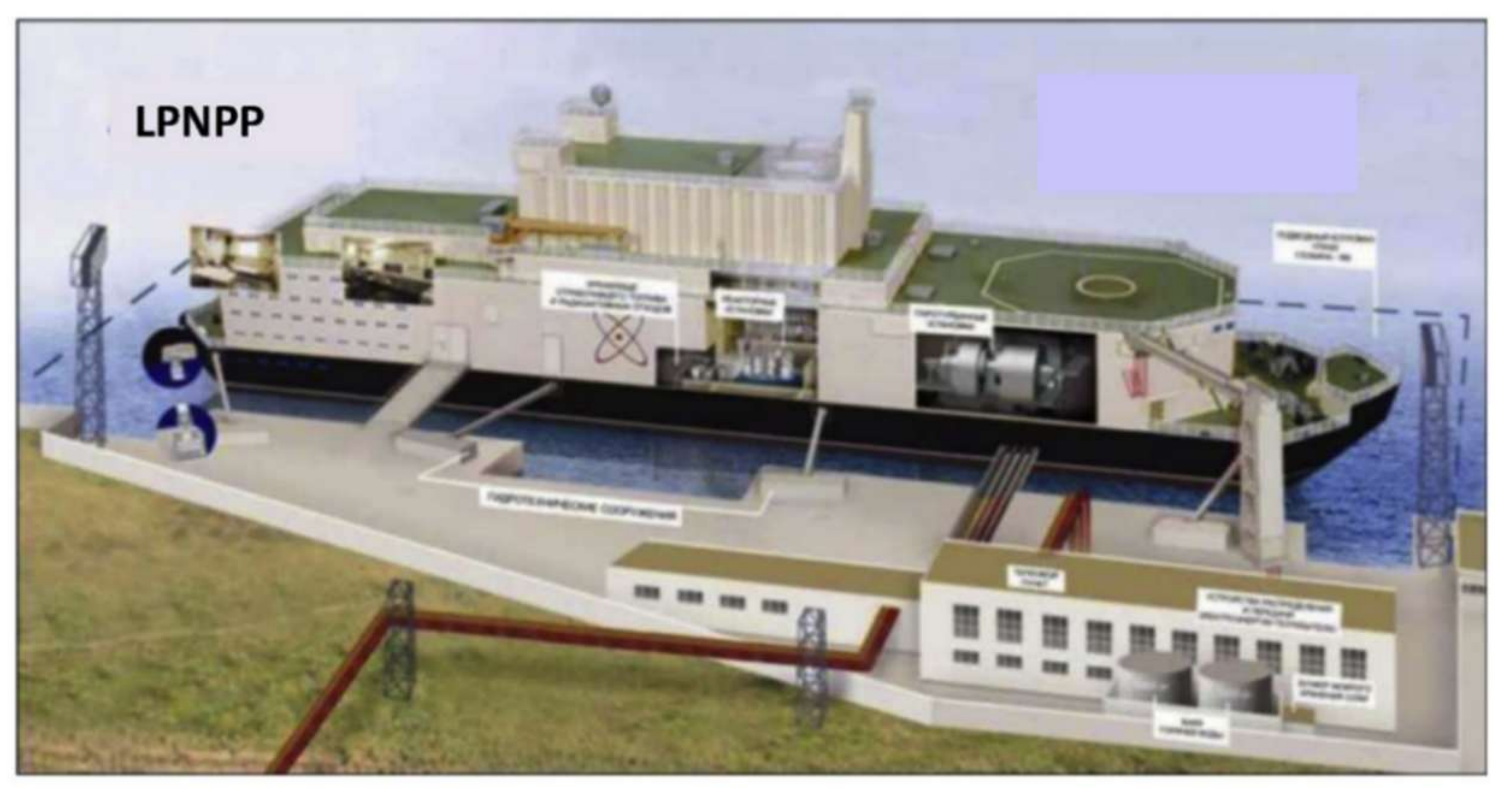

| | | Installed capacity | |
|---|---|---|---|
| Length, m | 140,0 | • electric, MW | 70/38 |
| Width, m | 30,0 | • thermal, Gcal/h | 50/146,8 |
| Hull height, m | 10,0 | Cost, RUB bn | 9 - 10 |
| Draft, m | 5,6 | construction time | 4 years |
| Tonnage, t | 21000 | plant output | |
| | | • electric supply, mln kWh/g | 465 |
| | | • heat supply, thous Gcal/g | 900 |

*Figure 6: Structure and features of a low-power nuclear power plant. Adapted from [33].*

### 4.11.4. Summary of Reviewed Papers

Table 4 summarizes the main findings on enabling technologies and emerging pathways.

*Table 4: Key findings on enabling technologies and emerging pathways.*

| Article | Summarized results | Year |
|---|---|---|
| Holdmann et al. [27] | Managing the high penetration of renewables requires flexible thermal loads, energy storage, and advanced grid-forming strategies within a supportive socio-technical framework to ensure reliable, resilient, and nearly 100% renewable operation. | 2019 |
| Das and Canizares [6] | Optimized integration of renewable energy with variable speed generators yields the best performance. | 2019 |
| Sambor et al. [104] | Investing in 17 kW of solar PV for an Arctic container farm can cut annual costs by up to 18%, even though it supplies only about 7% of winter and 25% of yearly load. | 2020 |
| Berger et al. [105] | Aggregating wind production sites across different continents may reduce the frequency of system-wide low-wind-power events. | 2020 |
| Dewitt et al. [15] | The results indicate that renewables can be a cost-competitive option and that the optimal mix of renewables varies for different communities. | 2020 |
| Elistratov et al. [30] | The study shows that an advanced Intelligent Automatic Control System for an Arctic wind-diesel hybrid system boosted renewable penetration, mitigated blade icing, and achieved up to 60% diesel savings and 20% OPEX reduction. | 2021 |
| Østegaard [106] | This review article addresses the status of research within the application of renewable energy sources. | 2022 |
| Temiz and Dincer [77] | An innovative ocean- and solar-based energy, food, fuel, and water production system is being developed for Arctic communities. | 2022 |
| Sambor et al. [26] | Adding 1.5 kW of solar PV with optimized Water Reuse system control can cut lifetime costs by over 13% and diesel microgrid bills by up to 37%, while using excess wind energy can reduce total project costs by 68%. | 2022 |
| Hunt et al. [90] | IDO-CAES is an innovative, low-cost, and environmentally friendly technology for long-term energy storage that uses isothermally compressed air in deep-sea or depleted gas reservoirs. | 2023 |

## 4.12. Governmental Subsidies and Legal Frameworks for Remote Arctic Communities

Government support in the form of public subsidies is essential to ensure the affordability of the electricity supply to consumers in remote areas [13]. A framework, on the other hand, is a set of rules and guidelines for developing an energy system. A functioning legal framework promotes a fair market environment for all energy sources. Remote Arctic communities often face significant environmental, social, and economic challenges due to the high cost and fragility of their energy supplies. Most goods and fuel are transported by sea when ocean access is available during the summer months, while air travel remains the primary mode of transportation in these regions [23]. These communities are often characterized by high electricity generation costs, mainly due to the transportation of diesel fuel to remote locations [21]. According to the IEA, consumers in remote areas *"might otherwise not be able to afford the 'true' energy costs of electricity generation."*. However, such subsidies can also undermine the competitiveness of renewable energy and reinforce reliance on fossil fuels in those regions [13].

High capital costs and technical challenges make renewable energy projects in remote areas difficult, and direct subsidies are often required to encourage investment. More generally, sectoral funds play an essential role in rural electrification (e.g., in Brazil). Based on these experiences, the IEA and the International Renewable Energy Agency recommend reorganizing energy subsidies, in particular by shifting fossil fuel subsidies to renewable energy subsidies [13]. In a study by Boute et al. [13], the authors showed that a structural change in energy subsidies in Russia's remote areas is needed to ensure the financial viability of investment in renewable energy systems. In a parallel study, Belonozhko et al. [107] suggest that Russia's energy sector needs strong economic (tax incentives) and legal support (favourable investment conditions) to boost competition and resolve regional energy supply issues. Holdmann et al. [28] showed that the absence of large electricity subsidies was a necessary condition for the development of community renewable energy projects.

The case of Alaska is exciting in this regard, where subsidies played a crucial role in increasing renewable energy penetration. Alaska was a pioneer in integrating renewable energy sources in the Arctic. It has one of the highest adoption rates of microgrid-enabling technologies in the world, accounting for approximately 12% of the world's total installed capacity. This push for renewable-powered microgrids is driven mainly by the goal of delivering affordable, reliable energy to remote communities, where extending traditional grid infrastructure is neither economical nor technically feasible [24].

As important as it is to develop such advanced energy systems that are future-proof for remote communities in the Arctic, Rodon et al. [108] argue that these development projects should not rely solely on technical and economic feasibility metrics but also consider post-colonial relations between indigenous communities and governmental institutions. They state that energy transition pathways are fundamentally social, they are "*woven into societal, geographic, and geopolitical arrangements at scales from the individual and the planet*". In their study, Gritsenko et al. [2] question the development of Arctic regions in Russia, which is underpinned by commercial resource exploitation. Resource colonialism is defined as economic activities, programs, discourses, and policies that promote extractive projects backed by central governments. While the local population has no representation in decision-making, it often faces a diverse and conflicting relationship between the state and the community. The study concludes that the Arctic's energy future reflects a tension between resource exploitation and environmental protection amid global market and policy shifts.

#### 4.12.1. Summary of Reviewed Papers

Table 5 summarizes the main findings on governmental subsidies and legal frameworks.

*Table 5: Key findings on government subsidies and legal framework studies.*

| Article | Summarized results | Year |
|---|---|---|
| Boute et al. [13] | A structural change to the subsidization of energy in Russia's remote areas is needed to ensure the financial viability of renewable energy investments. | 2016 |
| Bjørnebye et al. [41] | Differentiated feed-in premiums that reflect location-specific grid costs are essential to incentivize wind power development in socially optimal sites and minimize overall system costs, unlike uniform premiums, which lead to higher grid investment needs. | 2018 |
| Gritsenko et al. [2] | The Arctic's energy future reflects a tension between resource exploitation and environmental protection amid global market and policy shifts. | 2018 |
| Belonozhko et al. [107] | Russia's energy sector needs strong economic and legal support, including favourable investment conditions and tax incentives, to boost competition and resolve regional energy supply issues. | 2019 |
| Brazovskaia et al. [8] | The study confirms that increasing the share of renewable energy significantly reduces $CO_2$ emissions and supports the UN Sustainable Development Goals, making renewable energy systems essential for clean, affordable energy in remote regions like the Russian Arctic. | 2021 |
| Dewitt et al. [1] | This paper presents an overview of current electricity generation and consumption patterns in the Arctic. | 2021 |
| Dewitt et al. [10] | The results indicate that it is advisable to transition to renewable energy technologies rather than continue investing in diesel technology. | 2021 |
| Rodon et al. [108] | Distrust reflects a colonial mindset that questions indigenous capacity for self-determined development. | 2021 |
| Dewitt et al. [15] | Locally generated renewables can help to increase mid-term energy security by decreasing dependence on fossil fuels. | 2022 |
| Holdmann et al. [28] | The absence of large electricity subsidies was found to be a necessary condition for the development of community renewable energy. | 2022 |
| Pinto and Gates [32] | The subsidized price of diesel-based electricity and the lack of grants for renewable options provide residents (homeowners and renters) with little financial incentive for bottom-up change. | 2022 |

# 5. Design and Optimization of Hybrid Energy Systems in the Arctic

This section examines the studies concerning the planning, modelling and optimization of hybrid energy systems in the Arctic and other cold-climate regions. The studies analysed in this review have used different energy system models, including EnergyPLAN, TIMES, HOMER, TRNSYS, MATLAB, and a range of configurations of multi-integer linear programming MILP.

Energy models can be defined as "*a simplified description of reality with the goal of highlighting certain relations and to make the best prediction of future the developments*". These models can be considered a framework for the relations among the technical, economic, and social aspects of a particular energy system [31]. Similarly, the International Energy Agency defines energy models as *"abstraction of realities that simplify the world into bit-sized pieces in order to fit within certain sets of mathematical models"*. Energy modelling solves the "trilemma" of challenges related to energy systems, such as energy, environment, and economy, and helps the policy and decision makers to make well-informed decisions [31].

Energy system models support the assessment of future energy projects and facilitate the transition towards cleaner energy systems [19]. As energy system models are used to analyse the project feasibility of future installations, these models should be flexible, transparent, and balance general applicability with site-specific details to ensure higher levels of penetration and integration [23]. To achieve these models' specific characteristics, it is essential to understand the diversity of energy models and how different types serve different functions. The following sections offer an overview of these energy modelling systems. Several energy system models have been developed and applied. For further reading, see the reviews by Conolly et al. [43], Ringkjøb et al. [44], and Tozzi et al. [45].

## 5.1. EnergyPLAN

EnergyPLAN was developed at Aalborg University and has been continuously expanded since 1999 [43]. This modelling tool provides hourly estimates by simulating energy balances of all sectors for a year. It primarily focuses on integrating energy sectors and exploiting resources to enable flexible integration of renewable energy. Similar units are represented in aggregated form; for example, all wind turbines are represented by a single installed capacity. This modelling tool is designed to provide energy systems analysis for local (including municipalities), regional, national, and supranational scales [46].

Among the studies analysed in this review, EnergyPLAN was used in three articles. Galimova et al. [5] used the tool to simulate an energy system consisting of wind turbines, solar panels, and hydrogen (storage and e-fuel export) for renewable energy scenarios in Greenland in 2050. Sorknæs et al. [46] used EnergyPLAN to develop scenarios for transitioning from 3GDH ($3^{rd}$-generation district heating) to 4GDH ($4^{th}$-generation district heating). Similarly, Fischer et al. [47] coupled EnergyPLAN with a multi-objective optimization algorithm to identify optimal alternatives for Piteå energy system.

EnergyPLAN is deterministic and spatially aggregated. It normally analyses one year, and optimizes the operation of predefined systems rather than directly optimizing investment decisions [48].Although its hourly resolution captures variations within the selected year, its deterministic formulation does not inherently account for uncertainty or interannual variability. Its aggregated representation also limits the analysis of individual microgrid components. These limitation are particularly relevant to isolated Arctic systems, where accurately representing short-term renewable energy variability and periods of low solar and wind availability is essential for ensuring security of supply [19].

## 5.2. MARKAL/TIMES

This tool was developed by the International Energy Agency and provides system-level analysis of interaction among energy, the economy, and the environment (3E analysis). These are general-purpose model generators that take input data and simulate the evolution of the system over 20 to 100 years at global, multi-regional, national, state, or community scales. It can represent a wide range of thermal loads, renewable generation, energy storage systems, and transportation technologies [43]. Among the studies analysed in this review, Ringkjøb et al. [19] used this tool to analyse future energy systems for Longyearbyen. Bjørnebye et al. [41] used a special configuration (TIMES-Norway) to analyse the optimal locations of wind power plants in mainland Norway, while Astudillo et al. [49] used this tool to simulate the effects of heat pumps in building energy conservation in Canada.

When configured as a deterministic model with limited temporal resolution, TIMES can underestimate annual system costs, overestimate the contribution of variable renewable energy sources, and underestimate flexibility requirements. Although increasing temporal resolution improves model accuracy, it also substantially increases computational time [50]. This limitation is particularly relevant when assessing remote and isolated communities, where stochastic modelling is better suited for capturing operational uncertainty and variability in Arctic systems [19]. The trade-off becomes especially important in small Arctic systems, where renewable generation and energy demand can vary considerably. In the TIMES-Longyearbyen model, Ringkjøb et al. used 60 operational scenarios to represent stochastic parameters. Increasing the number of scenarios can improve the robustness of the results but also increases computational effort. Consequently, model transparency and validation are particularly important because results depend strongly on assumptions, temporal resolution and implementation choices by the modeler.

## 5.3. TRNSYS

TRNSYS is a transient systems simulation program developed and maintained by an international collaboration among the US, France, and Germany. It has an open, modular structure with open-source code that simulates the electricity and heat sectors of an energy system [43]. Multi-objective optimization tools are generally used to optimize energy models in TRNSYS [51]. In total, 10 studies used this software to simulate energy systems. Most studies have used this tool to analyse the thermo-environmental effects of energy systems. In two similar studies, Rehman et al. [52], [53] used it to test systems composed of heat pumps, storage tanks, and borehole thermal storage. Later, in subsequent studies by Rehman et al. [63], [65], [66], it was used to analyse solar, solar thermal, wind, battery storage, and seasonal thermal storage. Other studies used this tool to assess thermal energy demands [54], evaluating the performance of solar-assisted heat pumps SAHP [34], [55], [56], and analysing the geometry of borehole thermal energy storage systems, as in [3], where the authors showed the advantages of using a square-shaped BTES system.

The reliability of TRNSYS results depends on the suitability of its component models and the availability of representative local data. For example, Belzile et al. used weather data from a location approximately 160 km from the study site because local weather data was not available. The study also omitted some ground freezing effects and recommended further on site testing to validate the simulated energy savings [57]. Although TRNSYS supports detailed system simulation, optimization problems may require coupling with external optimization tools [58].

## 5.4. HOMER

HOMER, developed in 1992 by the National Renewable Energy Laboratory of the US, is a user-friendly micropower design tool that can be used to simulate and optimize stand-alone or grid-connected power systems that are composed of wind turbines, PV arrays, hydropower, biomass, ICE, microturbines, fuel cells, batteries, and hydrogen storage serving both electric and thermal loads [43]. Among the studies reviewed, Knowles et al. [11] used HOMER to optimize solar and wind-based energy systems with battery energy storage and sensible thermal energy storage. McKinley et al. [25] examined wind and solar generation combined with hydrogen storage. Chade et al. [59] and Kalantari et al. [14], [60] analysed systems with wind as the main source of energy, with diesel, battery storage, and hydrogen used as storage technologies.

In the Kotzebue study by Mckinley et al. [25], HOMER could not directly represent the community's bifacial PV arrays, and unplanned wind turbine downtime was represented by introducing a 24-hour shutdown period. The model was therefore validated against measured SCADA data, with simulated PV generation differing from the measured output by only 1.4% [25]. These findings indicate that HOMER is suitable for techno-economic microgrid analysis, although Arctic specific technologies and unplanned operational events may require approximations and validation using measured data [25].

## 5.5. MATLAB

MATLAB is an effective tool for modelling various renewable energy systems under varying constraints. The platform provides sufficient flexibility to analyse both the economic and operational performance of diversely configured energy systems [61]. Different configurations of multi-integer linear programming can be implemented on this platform. It offers advanced computational capabilities along with its ability to integrate control algorithms, like Model Predictive Control MPC. MPC is widely used for energy system optimization and management, including the coordination of energy production and consumption demands, by forecasting future demand and adjust system parameters to minimize cost and improve efficiency [62]. MATLAB is combined with compatible energy simulation and optimization platforms, for example, TRNSYS [58] and EnergyPLAN [63], in addition to providing several specialized tools (Simulink) designed for energy system planning [62]. Paulin et al. [39] used MATLAB to develop and evaluate novel microgrid configurations incorporating thermal storage units. Ninad et al. [21], Hosseini et al. [22] and Hyvonen et al. [61] used it as a tool to optimize various types of systems, including solar PV, wind, and diesel-hybrid configurations.

MATLAB/Simulink is a general-purpose modelling environment that provides a flexible environment for developing customized hybrid-energy-system models. In the thesis by Agrawal [64], the researcher developed a stand-alone model comprising diesel generators, batteries, PV arrays and wind turbines, together with economic and environmental analysis modules. The results were compared with those produced by HOMER and showed close agreement [64]. However, this flexibility requires users to define the system structure, component models, control or optimization methods, and economic calculations. Consequently, model transparency and validation are particularly important, as the results depend strongly on the assumptions and implementation choices made by the modeler.

## 5.6. Summary of Reviewed Papers

Table 6 summarizes the main findings of the studies reviewed in this section.

*Table 6: Summary of the main findings from studies on energy-system modelling and optimization.*

| Article | Summarized results | Year |
|---|---|---|
| Romero et al. [65] | This study optimizes energy supply for a remote Arctic mine, showing that adding two wind turbines saves 3% diesel while minimizing total energy costs. | 2016 |
| Quitoras et al. [31] | The optimal hybrid system proposed in this study saves 353,407 Liters of diesel annually, displaces 675 MWh/year, achieves over 70% renewable penetration, and reduces government electricity subsidies by 70%. | 2017 |
| Rehman et al. [51] | This study optimizes a centralized solar district heating network integrated with renewable electricity, showing that wind turbines and stationary batteries are crucial for Nordic conditions, while PV becomes more important with higher EV penetration. | 2019 |
| Fischer et al. [47] | The study optimizes Piteå's energy system using a multi-objective algorithm, reducing peak electricity imports by 38%, recommending heat pumps and biomass boilers for integrated heating-electricity systems. | 2020 |
| Ninad et al. [21] | PV integration in diesel-based microgrids is constrained by the generator's minimum loading (10%), allowing up to 45% PV capacity to supply about 10% of annual energy without curtailment. | 2020 |
| Ringkjøb et al. [19] | Transitioning remote Arctic settlements to renewable energy can be achieved through the integration of wind and solar power with hydrogen energy storage. | 2020 |
| Farrokhi et al. [56] | Dynamic simulation of the integrated CCHP system met all energy demands of a hotel, sold surplus power to the grid for 65% of the year, and achieved peak efficiency (37%) with the lowest GHG emissions. | 2021 |
| Nguyen and Bostrom [66] | A hybrid system of wind, solar PV with two-axis tracking, and a 3-kWh battery for a Tromsø household can achieve 98.2% self-sufficiency. | 2021 |
| Pantaleo et al. [67] | The study identifies solar PV with battery storage and diesel backup as the most viable near-term solution, while indicating that hydrogen remains too costly for short- to mid-term energy planning. | 2022 |
| Dumas et al. [68] | Optimizing a semi-detached house using genetic algorithms shows that solar PV systems can supply about one-third of the electricity needs. | 2024 |
| Hosseinnia et al. [69] | The most cost-effective and reliable solution for renewable energy integration in a cold-climate wastewater treatment plant WWTP is a holistic system combining solar PV, solar thermal, solar-assisted ground source heat pump SAGSHP, and lithium-ion batteries. | 2025 |

| Article | Summarized results | Year |
|---|---|---|
| Hosseini et al. [22] | Hybrid systems integrating wind, solar, and diesel cut emissions by over 55% and remain far more cost-effective than fully renewable setups, which are hindered by high capital and battery costs. | 2025 |
| Mousa et al. [70] | This study proposes a coordinated planning approach integrating renewables, EVs, batteries, heat storage, and dynamic demand response to achieve major improvements in voltage balance and hosting capacity, while recommending hydrogen storage to address seasonal mismatches. | 2025 |
| Knowles et al. [71] | The study evaluates the benefits of integrating battery energy storage with sensible thermal energy storage and finds that the combined system reduces costs across all renewable energy fractions. | 2025 |
| Hu et al. [72] | Replacing coal heating in rural building renovations cuts $CO_2$ emissions by 51–90%, with biomass pellet boilers achieving the lowest emissions, while PV-integrated heat pump systems offered the best balance of cost-effectiveness and environmental impact. | 2025 |
| Punyam and Gebremedhin [73] | This study introduces a planning framework for isolated microgrids using genetic algorithms, showing that renewable penetration beyond 70% sharply increases curtailment and costs, with battery capacity as a key limiting factor. | 2025 |
| Taghavifar et al. [74] | Two-tier optimization of an Arctic hybrid off-grid system integrating wave, wind, and hydrogen significantly reduces costs, aided by hydrokinetic turbines and optimized energy management. | 2026 |

# 6. Conclusion

The following conclusions are drawn from the survey results.

- Renewable energy penetration will increase in the remote Arctic and cold climate regions. However, a 100% renewable penetration is challenging without curtailment.
- In the renewable energy mix, the share of wind energy could be larger than that of solar energy because of low solar irradiance during the winter months and the year-round availability of wind resources.
- Storage plays a central role in remote cold-region Arctic energy systems. This review discusses three types of energy storage. Li-ion batteries for frequency and voltage regulation, thermal storage to cover heating demand, and hydrogen as a possible seasonal energy storage unit.
- Hydrogen was identified as a key component in many studies. The results suggests hydrogen as a storage material and as an energy carrier.
- Heat pump and borehole thermal storage unit were identified as key components of heating system in cold climate regions.
- Many remote Arctic communities rely on islanded grids; advanced intelligent control methods can optimize the overall system and reduce the required battery storage capacity.
- For modelling energy systems, this review suggests using specialized modelling tools that are specifically designed for arctic regions.
- Building retrofits can improve efficiency that is currently being lost due to old constructions that were built without building construction codes.
- Subsidies provided for diesel-based electricity generation should be redirected to support the development of new renewable energy systems. Until fully renewable and self-sufficient energy systems are developed, diesel can serve as an emergency backup.

**Acknowledgements:** The authors would like to acknowledge FME solar for funding this research. The authors also extend their sincere thanks to Mari Øgaard IFE for her valuable discussion, expert feedback and thoughtful suggestions, which contributed significantly to the development of this work.

# Appendix

## List of Abbreviations

| **Abbreviation** | **Full description** |
|---|---|
| 3GDH | 3rd-generation district heating |
| 4GDH | 4th-generation district heating |
| ASHP | Air source heat pump |
| BHE | Borehole heat exchanger |
| BTES | Borehole thermal energy storage |
| CCHP | Combined cooling, heat, and power |
| CHP | Cogeneration of heat and power |
| COP | Coefficient of performance |
| CSP | Concentrated solar plant |
| DH | District heating |
| DHW | Domestic hot water |
| DRO | Demand response optimization |
| DSM | Demand side management |
| DSO | Demand side optimization |
| DSR | Demand side response |
| DX-SAHP | Direct expansion solar-assisted heat pump |
| EV | Electric vehicle |
| FEC | Final energy consumption |
| GHG | Greenhouse gas |
| GSHP | Ground source heat pump |
| HP | Heat pump |
| IACS | Intelligent automatic control systems |
| IDX-SAHP | Indirect expansion solar-assisted heat pump |
| LCOE | Levelized cost of electricity |
| LSPV | Large-scale PV |
| MILP | Multi-integer linear programming |
| MPC | Model Predictive Control MPC |
| NWT | Northwest territories |
| NZEB | Net-zero emission building |
| PEB | Positive energy building |
| PEMFC | Proton exchange fuel cell |
| PTC | Parabolic trough collector |
| PV | Photovoltaic |
| RES | Renewable energy system |
| SAHP | Solar-assisted heat pumps |
| SMR | Small modular reactor |
| SOFC | Solid-state oxide fuel cell |
| TES | Thermal Energy Storage |
| WWTP | Wastewater treatment plant |

*Table 7: For Figure 1.*

| Publication year | Number of articles |
|---|---|
| 2015 | 1 |
| 2016 | 4 |
| 2017 | 2 |
| 2018 | 6 |
| 2019 | 8 |
| 2020 | 10 |
| 2021 | 12 |
| 2022 | 13 |
| 2023 | 3 |
| 2024 | 13 |
| 2025 | 14 |
| 2026 | 2 |

*Table 8: For Figure 2*

| Region | Number of articles |
|---|---|
| Canada | 20<br>[3], [6], [14], [21], [22], [23], [31], [39], [49], [54], [60], [65], [68], [69], [71], [77], [101], [108], [109], [110] |
| Russia | 13<br>[2], [4], [8], [12], [13], [29], [30], [33], [35], [76], [80], [81], [107] |
| Norway | 10<br>[19], [41], [42], [66], [73], [79], [99], [102], [111], [112] |
| Finland | 10<br>[34], [38], [52], [61], [78], [88], [95], [97], [98], [100] |
| USA | 7<br>[9], [24], [25], [26], [28], [103], [104] |
| Greenland | 3<br>[5], [67], [105] |
| Sweden | 2<br>[92], [113] |
| China | 2<br>[72], [91] |
| Iran | 2<br>[56], [94] |
| Japan | 1<br>[84] |
| Denmark | 1<br>[46] |
| Iceland | 1<br>[59] |
| Italy | 1<br>[55] |
| Germany | 1<br>[85] |
| Others | 14<br>[1], [10], [15], [16], [70], [75], [87], [90], [96], [106], [114], [115], [116], [117] |

*Table 9: Indicators for evaluation criteria of nZEB. Adapted from [99].*

| Category | Indicators |
|---|---|
| **Energy Self-Sufficiency** | Load Matching Index; Energy Saving Ratio; Design Mismatch Ratio; NZE Balance; |
| **Environment** | Carbon Dioxide Equivalent ($CO_2eq$) Emissions; Global Warming Potential (GWP) |
| **Economy** | Levelized Cost of Energy (LCOE); Life Cycle Cost (LCC); Net Present Value (NPV); Investment Costs; Operational Costs; Energy Payback Period |
| **Grid Stress Impact** | Grid Interaction Index; share of renewable energy (RE ratio) in grid; Energy Transmitted from Grid to ZEB; Transmission Losses |
| **Others** | Heat Pump Coefficient of Performance (COP); Exergy Efficiency; Solar Thermal Fraction; Solar System Efficiency; Power Losses |

*Table 10: An overview of the models used in the studies and specific optimizations performed.*

| Model | Articles | Optimization type |
|---|---|---|
| **HOMER** | [14], [25], [59], [71], [118] | A micropower design tool that can be used to simulate and optimize stand-alone or grid-connected power systems. |
| **TRNSYS** | [3], [34], [51], [52], [54], [55], [56], [95], [97], [100] | Tool to analyse the thermos-environmental effects of energy systems. |
| **EnergyPLAN** | [5], [46], [47] | Integrating energy sectors and exploiting sources to enable flexible integration of renewable energy. |
| **TIMES** | [19], [41], [49] | System-level analysis of energy, the economy, and the environment (3E analysis). |
| **MATLAB** | [21], [39], [61] | An effective tool for modelling various renewable energy systems. |